\documentclass[sigconf]{acmart}
\acmConference[ESEC/FSE 2022]{Submitted ACM Joint to the European Software Engineering Conference and Symposium on the Foundations of Software Engineering}{14 - 18 November, 2022}{Singapore}
\AtBeginDocument{%
  \providecommand\BibTeX{{%
    \normalfont B\kern-0.5em{\scshape i\kern-0.25em b}\kern-0.8em\TeX}}}
\setcopyright{acmcopyright}
\copyrightyear{2022}
\acmYear{2022}
\acmDOI{XXXXXXX.XXXXXXX}
\acmPrice{15.00}
\acmISBN{978-1-4503-XXXX-X/18/06}

\usepackage{enumitem}
\usepackage{hyperref}
\usepackage{listings}
\usepackage{color}

\definecolor{mygreen}{rgb}{0,0.6,0}
\definecolor{mygray}{rgb}{0.5,0.5,0.5}
\definecolor{mymauve}{rgb}{0.58,0,0.82}

\definecolor{chestnut}{rgb}{0.8, 0.36, 0.36}

\begin{document}
\begin{sloppy}

\title{In War and Peace:\\The Impact of World Politics on Software Ecosystems}

\author{Raula Gaikovina Kula}
\affiliation{%
  \institution{Nara Institute of Science and Technology, Japan}
  \country{}
}
\email{raula-k@is.naist.jp}

\author{Christoph Treude}
\affiliation{%
  \institution{University of Melbourne, Australia}
  \country{}
}
\email{christoph.treude@unimelb.edu.au}

\begin{abstract}
Reliance on third-party libraries is now commonplace in contemporary software engineering.
Being open source in nature, these libraries should advocate for a world where the freedoms and opportunities of open source software can be enjoyed by all.
Yet, there is a growing concern related to maintainers using their influence to make political stances (i.e., referred to as protestware).
In this paper, we reflect on the impact of world politics on software ecosystems, especially in the context of the ongoing War in Ukraine.
We show three cases where world politics has had an impact on a software ecosystem, and how these incidents may result in either benign or malignant consequences.
We further point to specific opportunities for research, and conclude with a research agenda with ten research questions to guide future research directions. 
\end{abstract}


\maketitle

\section{Introduction}

Contemporary software is not built in isolation.
With the emergence of third-party libraries, and massive software repositories, now more than ever developers are able to quickly adopt functionality into their applications, avoiding the time-consuming and error-prone task of writing and testing from scratch.

The success of these libraries is made possible through the concept of Open Source Software (OSS).
OSS allows software to be free for anyone to use, which is incorporated into everything from independent projects to mainstream, proprietary consumer software. 
Being OSS, libraries are typically maintained by one or a handful of volunteers, and they benefit from having further volunteer contributors, as generally the developers in a software ecosystem should be able to help each other by pointing out issues and contributing new features.
Libraries harness the power of distributed peer review and transparency of process, promising higher quality, better reliability, greater flexibility, and lower cost.
Projects build reputation over time, with developers gaining trust in using these libraries. 
Each ecosystem has their own culture of maintainers, who are empowered to approve and publish contributed code changes.
Evidence of the impact of libraries is seen with the growing size of NPM with more than one million packages~\cite{mujahid2022characteristics}, for example, and the growing support for these ecosystems, e.g., GitHub's 2020 acquisition of NPM.\footnote{\url{https://github.blog/2020-03-16-npm-is-joining-github/}}

Problems arise when a maintainer feels empowered to sabotage their own projects, thus weaponizing their library as protestware, e.g., with the intention to make users of their library aware of some political stance, or situation.
% \raula{
% There's no way to completely remove the threat that a maintainer of an open source project will go rogue, either for personal reasons or because of a criminal or government influence. But so-called “insider threats” can't completely be eliminated within private companies either. The open source community and major influences like Github are increasingly looking to automated code scanning tools to put more eyes (if digital ones) on even the most esoteric projects and catch more bugs or potentially suspicious changes before they go live or soon after.
% Casting such a wide net is particularly important because of another problem in open source security in which bad actors infiltrate projects or convince burned out maintainers to hand over the reins and then have full control to deploy whatever they want. Automated scanners have limitations, though, and Lorenc notes that they are often better at catching accidental bugs than those that are intentionally designed for sabotage.
% Longtime open source security researchers and practitioners are adamant, though, that another vital safeguard exists right out in the open: massively expanding the support and resources maintainers can seek in general and especially if their fun hobby project eventually morphs into a critical link in the global software supply chain.
% }
% \raula{
% This blog attempts to tie everything together. 
% \url{https://beny23.github.io/posts/on_weaponisation_of_open_source/}
% }
Responses from practitioners to such protestware have been mixed.
According to a developer blog that was later republished by the IEEE Computer Society~\cite{Web:Weaponize}, practitioners have expressed their concerns, stating that \textit{``It's ill-considered and user-hostile, and can trivially go wrong. Weaponizing open source to inject malware, no matter how well intentioned, is still injecting malware''.}
Others question whether this is Open Source anymore, as a OSS licence clearly states that there should be ``\textit{No Discrimination Against Persons or Groups. The license must not discriminate against any person or group of persons}''~\cite{Web:OSI}.
The official statement from the Open Source Initiative (OSI) community lays the responsibility to maintainers, stating that ``\textit{Protest is an important element of free speech that should be protected. 
Openness and inclusivity are cornerstones of the culture of open source, and the tools of open source communities are designed for global access and participation. Instead of malware, there are so many outlets for open source communities to be creative without harming everyone who happens to load the update.
Longer term, the downsides of vandalizing open source projects far outweigh any possible benefit, and the blowback will ultimately damage the projects and contributors responsible. Use your power, yes—but use it wisely}''~\cite{web:OSIstate}.

The purpose of this article is to point the software engineering research community to open questions regarding how researchers can investigate, address, and regulate such kinds of protestware.
In light of the War in Ukraine, we present three motivating scenarios where world politics has had impact on software ecosystems, highlighting the side affects, and then present an agenda on how to dissect and respond to such behaviour during software engineering practices.
Another blog \cite{Web:blogSabotage} raises growing concerns within an ecosystem: ``\textit{Protestware can deliver similar anti-war messages, but within the open-source community there are worries that the possibility of sabotage — especially if it goes further than simple anti-invasion messaging and starts destroying data — can undermine the open-source ecosystem. Although it is less well known than commercial software, open-source software is enormously important to running every facet of the internet}''.
Our vision is that the lines of research outlined in this article can contribute towards building resilient and yet open software ecosystems.

In the remainder of the paper, we first introduce three motivating scenarios related to the War in Ukraine in Section~2 before we briefly discuss protestware in other contexts in Section~3. Section~4 presents the research agenda on dissecting the status quo of protestware, increasing open-source supply chain resilience, and managing trust and responsibiliy in software ecosystems, and we conclude the paper in Section~5.

% another case where GitHub is suspending accounts. 
% \url{https://www.pcmag.com/news/github-reportedly-suspends-accounts-related-to-sanctioned-russian-orgs}
% here is another tie \url{https://www.bleepingcomputer.com/news/security/third-npm-protestware-event-source-polyfill-calls-russia-out/}
% \url{https://www.bleepingcomputer.com/news/security/big-sabotage-famous-npm-package-deletes-files-to-protest-ukraine-war/
% \url{https://gist.github.com/MidSpike/f7ae3457420af78a54b38a31cc0c809c}

%  another one \url{https://www.jessesquires.com/blog/2022/04/19/github-suspending-russian-accounts/}
% importance of software ecosystems
% world events can interfere with SE, e.g., covid
% \section{Background and Related Work}
% Related work on issues in ecosystems, e.g., bugs/vulnerabilities
% Related work on the intersection of politics and SE
% other related work?
% \raula{
% Action by MongoDB
% https://www.theregister.com/2022/03/15/mongodb_terminates_russian_saas/
% }

\section{Motivating Cases}
In this section, we discuss three cases where protestware has had an impact on a software ecosystem. 
Specifically, we focus on the War in Ukraine, and three different effects that resulted from maintainers making a political stance.
The first case involves the malignant effects of protestware.
The second case is a benign case where protestware had no malicious intent to harm the users of a library.
In the last case, we take a look at a case where sanctions were placed on accounts that were related to world politics.

% Concrete examples described in detail, similar to Section 2 here: \url{https://ctreude.files.wordpress.com/2020/03/icse20b.pdf} Could use more than three!
% https://www.wired.com/story/open-source-sabotage-protestware/
% all about protestware.
% https://opensource.org/newsletter/OSI-mar-2022
% this is the Official statement
% https://opensource.org/blog/open-source-protestware-harms-open-source

\subsection{Case 1: Malignant Protestware}
% https://snyk.io/blog/peacenotwar-malicious-npm-node-ipc-package-vulnerability/
% https://github.com/RIAEvangelist/node-ipc/issues/233#issuecomment-1068182278
A developer of the JavaScript library node-ipc~\cite{Web:node-ipc}, which is used by the popular vue.js framework, deliberately introduced a critical security vulnerability that, for some netizens, would destroy their computers' files.
The library is fetched about a million times a week from the NPM registry, and is described as an ``inter-process communication module for Node, supporting Unix sockets, TCP, TLS, and UDP''.
Seemingly, the maintainer intentionally changed his code to overwrite the host system's data, then changed the code to display a message calling for world peace, as a protest against Russia's invasion of Ukraine.
GitHub declared this a critical vulnerability, which was tracked as CVE-2022-23812~\cite{Web:CVE-2022-23812}.

The malicious code was intended to overwrite arbitrary files dependent upon the geo-location of the user's IP address, attacking software in specific locations.
Concretely, the affected versions 10.1.1 and 10.1.2 of the library check whether the host machine has an IP address in Russia or Belarus, and if so overwrites every file it could with a heart symbol. 
Version 10.1.3 was released soon after without this destructive functionality, while 10.1.1 and 10.1.2 were removed from the NPM registry.

There was a strong response from the community, including frustrations that led to insightful discussions.
One example from a contributor on the GitHub Discussions channel is shown below~\cite{Web:discussion}:
\begin{quote}
\textit{I'm very happy to see that the principles and character of many in tech (FOSS especially) remain clear enough to recognize how completely wrong this was. Of course, if the marketplace of current things keeps hammering away at this, it will benefit a small number of corporate giants (misplaced trust/safety). I hope we all start seeing these patterns as we grapple with a general blurring of lines between tools for marketing and weaponry.
It's essential to ask: what's the outcome and who benefits? I like to ask the faux ideologues ``who agrees with you?'' ``Isn't it strange how well aligned you are with a small number of very visible, influential, and powerful organizations?'' ``What's the fight and who is on which side, again?''
It's about competency, not power. Power feeds and is fueled by egocentrism (plainly, weak vanity). Competency comes from discovering your natural gifts and applying them.}
\end{quote}

Another user from that Github Discussion quoted how this affected the Open Source Community \cite{Web:discussion}:
\begin{quote}
 \textit{The trust factor of open source, which was based on goodwill of the developers is now practically gone, and now, more and more people are realizing that one day, their library/application can possibly be exploited to do/say whatever some random dev on the internet thought was `the right thing to do'.}
\end{quote}
The maintainer defended his module on GitHub, saying ``this is all public, documented, licensed and open source''. Earlier, there were more than 20 issues flagged against node-ipc about its behavior.
Some of the comments referred to the creation as \textit{``protestware, while others might call it malware''}. 

\subsection{Case 2: Benign Protestware}
% Removing or lock npm supply chains of dependencies.
% https://www.theregister.com/2022/03/18/protestware_javascript_node_ipc/
% https://github.com/RIAEvangelist/peacenotwar/blob/main/WITH-LOVE-FROM-AMERICA.txt
We present two cases where the protestware does not have malicious intent, but aims at increasing awareness.
For the first case, the same maintainer of the node-ipc library then created the peacenotwar library~\cite{Web:node-peacenotwar}.
As explained by the maintainer, it serves as a non-violent protest against Russia's aggression.
Instead of malicious deletion of files, the module adds a message of peace on users' desktops~\cite{Web:peacecommit}.
The maintainer was quoted in the README file\footnote{\url{https://github.com/RIAEvangelist/peacenotwar}}:
\begin{quote}
    \textit{I pledge that this module, to the best of my knowledge and skills, does not do any damage to anyone's data. If you do not like what this module does, please just lock your dependencies to any of my work or other's which includes this module, to a version you have code reviewed and deemed acceptable for your needs. Also, please code-review your other modules for vulnerabilities.}
\end{quote}

When accepted, this code was included:

\begin{lstlisting}[frame=single]  % Start your code-block
    
    variable "putin_khuylo"{
    description = "Do you agree that Putin doesn't respect Ukrainian sovereignty and territorial integrity? More info: https://en.wikipedia.org/wiki/Putin_khuylo!"
    type        = bool
    default     = true}
\end{lstlisting}
Another example is given by a maintainer of the terraform modules for AWS who added their own protest in the licence file, with ``Additional terms of use for users from Russia and Belarus''~\cite{Web:terraform}:

\begin{quote}
    \textit{By using the code provided in this repository you agree with the following: \\
    - Russia has illegally annexed Crimea in 2014 and brought the war in Donbas followed by full-scale invasion of Ukraine in 2022.\\
    - Russia has brought sorrow and devastations to millions of Ukrainians, killed hundreds of innocent people, damaged thousands of buildings, and forced several million people to flee.}
\end{quote}

These two libraries are examples of protestware that are not intended to be malignant, compared to the case described in the previous section. 

\subsection{Case 3: Developer Sanctions}
The third case is not related to a single library maintainer, but affects a software ecosystem more broadly.
We report two instances.
The first is the decision of MongoDB not to sell its products to Russian buyers~\cite{Web:mongodb}, using the following statement:
\begin{quote}
    \textit{The breadth of the new sanctions from the US and internationally is unprecedented, and MongoDB has taken action to comply with them. We will not sell our cloud services to customers in Russia and Belarus and we will not sell any more MongoDB software to customers in Russia or Belarus.}
\end{quote}

MongoDB's decision appears to hinge on an interpretation of Western sanctions that say SaaS subscriptions represent a sale. In order to comply with sanctions, MongoDB decided it should not continue to offer its wares, not even as a service.
Interestingly, Oracle also stated on Twitter\footnote{\url{https://tinyurl.com/8admb86d}} that it is ending support for Russian users,
Even Microsoft joined Apple and SAP in suspending sales in Russia.\footnote{\url{https://www.siliconrepublic.com/enterprise/microsoft-sales-russia-ukraine-cyberattacks}}

In another instance of developer sanctions, at first GitHub suspended Russian accounts.
According to a blog post from a developer~\cite{Web:sanction}, ``suspending an account'' on GitHub meant deleting all activity for a user---which results in (1) every pull request from the suspended account being deleted, (2) every issue opened by the suspended account being deleted, and (3) every comment or discussion from the suspended account being deleted. In effect, the user's entire activity and history evaporated. 
The bigger issue was that GitHub had not provided any warning to the library maintainers.
As explained by a maintainer in a blog post~\cite{Web:sanction}:
\begin{quote}
    \textit{I recently took over as a lead maintainer for two popular projects in the Apple developer community, Quick and Nimble. I just released version 5.0 of Quick a few days ago. During the week leading up to the release, I was reviewing and merging many pull requests. But when it came time to write the release notes, I noticed very bizarre behavior. Mysteriously, some pull requests were deleted. Poof. Gone. Then I realized that an entire contributor's presence had disappeared — all of their comments on issues were missing, all of the issues they opened were gone, all of the pull requests they opened had vanished. Every piece of activity related to the user was gone.}
\end{quote}

GitHub later reached out to this maintainer, letting them know that it had restored the missing pull requests, issues, comments, etc.~from the Russian developers whose accounts had been suspended. 
User profiles were also restored, although GitHub did not specifically mention that the accounts had been suspended~\cite{Web:githubmsgukraine}.

% \raula{
% https://web.archive.org/web/20210704022108/
%https://github.com/Marak/faker.js/issues/1046}
\section{Other Maintainer Stances}
The War in Ukraine is not the first time that open-source maintainers have used their open source libraries as a platform for protest.
For example, Faker.js\footnote{\url{https://fakerjs.dev/}} and Colors.js\footnote{\url{https://colorjs.io/}} created problems for users of Amazon's Cloud Development Kit. 
Big companies, critics have long said, benefit from open source ecosystems without adequately compensating developers for their time. In turn, developers responsible for the software are unfairly strained.
The maintainer of these two JavaScript libraries with more than 21,000 dependent apps and more than 22 million weekly downloads, intentionally released an update that produced an infinite loop that caused dependent apps to spew gibberish, prefaced by the words ``Liberty Liberty Liberty''. 
His stance was made clear in the README file:
``Take this as an opportunity to send me a six-figure yearly contract or fork the project and have someone else work on it''~\cite{Web:fakermsg}.

These cases show that protestware is not only relevant in the context of political conflicts, but has a long history of capturing and communicating the opinions of open source maintainers.

\section{Research Agenda}
In this section, we present our research agenda.
We have presented three motivating scenarios where world politics has had an impact on a software ecosystem.
These cases form initial evidence on the current state of practice, and how the nature of open source software and protestware will affect the software supply chain, trust, and resilience within an ecosystem. 
In this context, a software supply chain attack occurs when a compromised library distributes malicious code to applications that depend on it.\footnote{\url{https://capec.mitre.org/data/definitions/437.html}} 

As discussed in Section 3, there is anecdotal evidence from blogs and developer discussions on the topic, yet there have been very few research studies that articulate the impact of protestware on software engineering.
From these motivating cases, the following questions about the possible implications have emerged.
We discuss each of these potential implications below in turn, and suggest ten research questions that could form the basis of future research projects.

\subsection{Dissecting the status quo of protestware}

Evidence from this paper shows the thin line that exists between protestware and malware.
As mentioned by the OSI community, protest is an important element of free speech, with openness and inclusivity being cornerstones of the culture of open source.
However, vandalizing open source projects threatens any possible benefit, and might damage the projects and contributors responsible.
Our first four research questions target the potential of protestware.

\begin{enumerate}
    \item How effective is protestware at communicating political messages?
    \item What is the immediate impact of protestware on a software ecosystem?
    \item Who is affected by protestware, and what relationship do they have with the protestware maintainer, if any?
    \item What mitigation strategies do projects use for securing or repairing their supply chains, and how effective are those mitigation strategies?
\end{enumerate}

Through in-depth studies of cases such as the ones introduced in Section 2, we propose research questions aimed at understanding the status quo of protestware, to build the foundations for further work on increasing resilience and trust in software ecosystems. We propose to survey developers that rely on the modified projects to determine whether protestware managed to achieve its goals of political messaging. Maintainers interested in taking political action are confronted with dilemmas of instrumental vs.~value rationality, i.e., selecting the most effective and efficient means to reach given ends (instrumental rationality) or acting without considering the foreseeable consequences in the service of conviction (value rationality)~\cite{rutgers2006morality, oakes2003max}. Is the creation of protestware a legitimate means to pursue a political agenda? Answers to such questions lie beyond traditional software engineering research, but we can investigate whether the message is received as intended, and what its immediate consequences are. Who is affected? A small number of developers who have a working relationship with the acting maintainer (e.g., through reciprocal contributions to each others' code), or an entire ecosystem? Using case studies, we can further investigate which strategies other players in a software ecosystem have employed to work around protestware and/or mitigate its impact, with the ultimate goal of learning how to increase resilience in the future.

\subsection{Increasing supply chain resilience}
In response to the attacks on supply chains, the Linux Foundation's partner group -- Open Source Security Foundation (OpenSSF), Google, and Microsoft joined forces to work with security experts and use automated security testing to improve open-source security in a project called the Alpha-Omega Project~\cite{Web:alphaomega}. 
This is a global effort to secure code, and has sparked other efforts such as weak links analysis \cite{Weak2022}.
Hence, our next three research questions are related to the supply chain.

\begin{enumerate}[resume]
    \item How effective are redundancies in supply chains at increasing resilience?
    \item How do changes which turn libraries into protestware differ from other software evolution?
    \item How accurately can we automatically detect protestware?
\end{enumerate}

Efforts to increase supply chain resilience could form a spectrum from manual efforts to minimise the impact of protestware by reducing the reliance on external libraries~\cite{xu2020reinventing} or introducing redundancies into supply chains~\cite{gherbi2011software} to automated tools for protestware detection and mitigation, similar to vulnerability detection~\cite{li2016vulpecker}. To pave the way for automated detection and mitigation, we propose research questions related to unique characteristics of patches which turn a library into protestware (e.g., are such patches `surprising' to a language model trained on past patches?~\cite{caddy2022surprisal}) as well as automated prediction, borrowing methods from the defect prediction literature~\cite{wan2018perceptions}. 

\subsection{Managing trust and responsibility}

Prior work has established that the success of a library is based on the library itself, which involves the assumption of a module's functional and non-functional correctness. For instance, system maintainers need to trust the reliability of non-functional attributes such as security and stability of an adopted library~\cite{KulaSANER2015}.
This trust in components is well-known in other fields, such as Dependable and Secure Computing~\cite{Guha2004, Hasselbring2006, Yan2011}.
However, we argue that trust also falls back on the maintainer. 
The final three research questions are aimed at maintainers and their role in an ecosystem as a whole.

\begin{enumerate}[resume]
    \item What are the responsibilities of maintainers, as perceived by other stakeholders in a software ecosystem?
    \item How does protestware affect trust into a library and entire ecosystems?
    \item How is protestware regulated at ecosystem level?
\end{enumerate}

Recent anecdotes, such as a Fortune-500 company unabashedly requesting a curl maintainer's immediate response about the Log4J vulnerability~\cite{Web:Log4j}, highlight the current confusion about what exactly are the responsibilities of a great open-source maintainer~\cite{dias2021makes}. The recent emergence of protestware only adds to this confusion---do maintainers act irresponsibly if they use open source for political action? We propose to capture the perceptions of different players in a software ecosystem about these responsibilities and how protestware can affect trust in single libraries or entire ecosystems.
It is important that software ecosystems are resilient to threats against their culture, thus becoming sustainable~\cite{kula2019life}. 
As mentioned in Section 1, open source licences contain statements about discrimination~\cite{Web:OSI}, but they are not explicit about protestware. We also propose to investigate how codes of conduct~\cite{tourani2017code} at project and ecosystem level can play a role in regulating the management of protestware.

\section{Discussion and Conclusion}

Protests are a powerful way for people to make their voices heard. They can be used to call attention to injustices, and to demand change. In the world of open source software, protests take the form of ``protestware''. This is software that is released in protest against a particular decision or action by an open source project. In highly inter-connected and large software ecosystems, where the average package can directly depend on more than five other packages (NPM~\cite{kula2017impact}), the impact of protestware on the entire ecosystem can be devastating, especially if it is malignant in nature. Using the context of the ongoing War in Ukraine, we have argued how and why protestware does have an impact on software ecosystems, and we have outlined a comprehensive research agenda for understanding and addressing protestware and its implications on ecosystem resilience, trust, and responsibility.

To answer the research questions posed above, we need a systematic research approach.
To confirm the anecdotal evidence, we need methods to define and detect different forms of protestware.
Perceptions of developers as well as viewpoints of large Fortune-500 companies would need to be sought, e.g., via surveys or interviews.
Furthermore, political stances can become a sensitive topic, so care needs to be taken in the design on how to effectively gather and interpret our results. 

Our hope is that answering these questions will help us understand how to sustain and build resilient software ecosystems.  

\section*{Acknowledgement}
This work is supported by Japanese Society for the Promotion of Science (JSPS) KAKENHI Grant Numbers 20K19774 and 20H05706.

% Subsections for different research topics, each with ideas for research questions, similar to Section 4 here: \url{https://ctreude.files.wordpress.com/2016/01/foser10.pdf}

% \raula{https://publications.teamusec.de/2022-oakland-sec-oss/pdf/committed-to-trust-preprint.pdf}

% \raula{
% https://sel.ics.es.osaka-u.ac.jp/lab-db/betuzuri/archive/979/979.pdf
% }

\bibliographystyle{ACM-Reference-Format}
\bibliography{warandpeace}

\end{sloppy}
\end{document}